# Experimental and numerical investigation of photoacoustic resonator for solid samples

S. El-Busaidy[1,2], B. Baumann[1], M. Wolff[1], L. Duggen[2], H. Bruhns[1]
1. Department of Mechanical and Production Engineering, Hamburg University of Applied Sciences, Hamburg, Germany
2. Mads Clausen Institute, University of Southern Denmark, Sønderborg, Denmark

**Abstract**

The photoacoustic signal in a closed T-cell resonator is generated and measured using laser based photoacoustic spectroscopy. The signal is modelled using the amplitude mode expansion method, which is based on eigenmode expansion and introduction of losses in form of loss factors. The measurement reproduced almost all the calculated resonances from the numerical models with fairly good agreement. The cause of the differences between the measured and the simulated resonances are explained. In addition, the amplitude mode expansion simulation model is established as a quicker and computationally less demanding photoacoustic simulation alternative to the viscothermal model. The resonance frequencies obtained from the two models deviate by less than 1.8%. It was noted that the relative height of the amplitudes of the two models depended on the location of the antinodes within the resonator.

**1. Introduction**

The interest in photoacoustic spectroscopy (PAS) has increased with the development of new high power infra-red laser sources [1]. The technique is highly sensitive, enabling measurement of weakly absorbing and optically opaque samples that cannot be measured using transmission spectroscopy [2,3]. The photoacoustic (PA) effect has already demonstrated its potential for various applications for both gaseous and solid samples [3,4,5]. This has resulted in the proposal of numerous photoacoustic-based sensors.

PA sensors typically use infrared laser radiation to excite vibrational states of molecules. Thermal de-excitation leads to a local temperature elevation. Since the radiation is modulated, the heat release by the molecules is as well. This generates periodic pressure changes in the surrounding environment that are detected as acoustic waves using a microphone or a tuning fork [6,7,8].

The signal produced by the PA effect is, however, usually weak and needs to be amplified. This can be achieved by a modulation of the electromagnetic radiation at an acoustic eigenfrequency of the resonator, thus exciting the corresponding acoustic mode. This significantly boosts the PA signal and increases the detection sensitivity of the sensor. The amplification of the PA signal can be maximized by optimizing the shape of the resonator. In a purely experimental investigation, various resonator shapes must be tested since the optimal geometry for maximum signal amplification is not obvious. This would be extremely time consuming and expensive. Therefore, numerical simulation methods are preferred [9,10,11].

The viscothermal (VT) model is considered the most accurate numerical method for simulating PA signals [10]. It can accurately map the loss effects at the surfaces of the resonator which are the dominant loss mechanisms in acoustic resonators. However, the VT model is computationally demanding requiring a lot of memory space and simulation time. In this article, we investigate the amplitude mode expansion (AME) model which is considered faster and computationally less demanding [11].

Glière et al. compared the results of the AME model to those of the VT model and a third approach in a micro-resonator by looking at a single resonance [10]. The comparison was carried out for a gaseous sample in a differential Helmholtz resonator. Baumann et al. applied the AME model to estimate the PA signal of nitric oxide molecules in an hourglass resonator [16]. They simulated a Gaussian shaped heat source across the resonator to mimic the absorption of the laser beam by the nitric oxide molecules across the resonator. Here we will extend the study of the AME model by simulating the PA signal of a macroscopic T-cell over a wide frequency range from 7 kHz to 62 kHz.

We are interested on the PA signal produced by a solid sample instead of a gaseous sample. PA signal measurement of the simulated resonator is carried out and compared to the simulated results.

## 2. Materials and method
### 2.1 Experimental setup

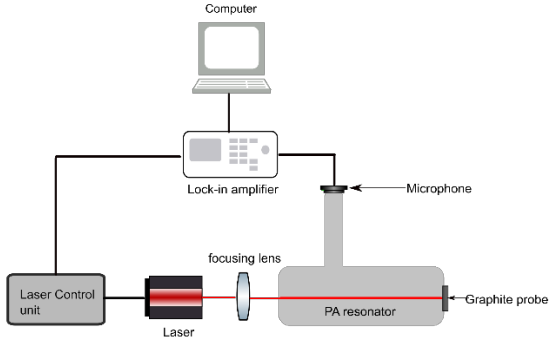

**Figure 1**: Schematic of the experimental setup

The setup for the photoacoustic signal generation is represented in Figure 1. A diode laser (IMM photonics U-LD-651071A) served as the radiation source producing optical power of approximately 200 µW at a wavelength of 655 nm. A signal generator (Agilent 33220A) is used to modulate the laser with 50% duty cycle. The beam is focused onto a graphite probe using a spherical lens (f=30 mm). Graphite is a broad band absorber in the visible region and can therefore be excited with the selected laser. The generated PA signal is detected using a digital ultrasound MEMS microphone (Knowles SPH0641LU4H-1). The microphone output is a digital pulse density modulated signal (PDM) and is passed through a low-pass filter circuit for demodulation. The demodulated signal is fed into the lock-in-amplifier (Signal Recovery7265 DSP Lock-in-Amplifier) to amplify and filter out noise from the generated PA signal. The measurements were done with a time constant of 100 ms. The modulation frequency was scanned between 8 kHz and 62 kHz at an increment of 10 Hz. For each frequency value, an average of 10 measurements was recorded with a computer. Two additional measurements were carried out between 8 kHz to 10 kHz. In one of the measurements, the graphite probe was replaced with glass window and in the other measurement, the laser was turned off to record the offset level.

### 2.1.1 T-cell resonator

The photoacoustic resonator consists of three interconnected cylinders as shown in Figure 2. A small *absorption cylinder* is longitudinally connected to a *cavity cylinder* which has a *resonance cylinder* perpendicularly mounted, thus forming a T-like structure. The right end of the resonator where the laser beam enters is sealed using a polycarbonate window with an optical transmission of 85% in the visible region. The graphite probe is pressed into the absorption cylinder thus sealing the left end of the resonator. The upper end of the resonance cylinder is sealed by the microphone and its mounting.

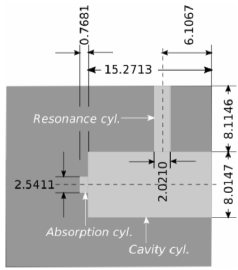

**Figure 2:** Cross-section of the T-cell resonator (light gray) showing the size of each cylinder in mm. The dimensions of the cells are the mean values obtained from a high precision measurement of the resonator [14].

The resonator was manufactured by drilling out the cylinders from an aluminum block. Its dimensions were measured to check for geometrical imperfections using an optical scanner (ZEISS MICURA) with a precision of less than 1 micrometer. Figure 3 shows that the diameter varies along each cylinder resulting in a rough interior surface. The deviations of the diameters are in the range of 0.01 mm to 0.06 mm and have been magnified by a factor of 2500 so that the unevenness of the cylinder is clearly visible.

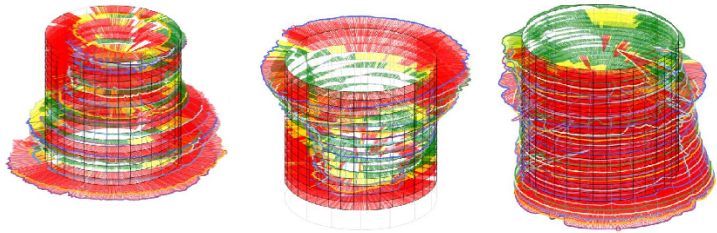

**Figures 3:** The measured dimensions of the absorption, resonance and cavity cylinders respectively (left to right). [14].

### 2.1.2 Microphone operation

The PDM digital interface of the used microphone is designed to allow time multiplexing of two microphone outputs on a single data line using a single clock. This is achieved by transmitting the output pulse train of one microphone in the first half and of the other microphone in the second half of the sampling period. However, only one microphone is used in the setup. Therefore, no output signal is issued during one half of the sampling period causing a discontinuous pulse train. For a low-pass filter used as PDM digital to analog converter (DAC), the analog output voltage is the product of DAC operating voltage and PDM duty cycle. For the maximum microphone output signal, the pulse duty cycle is 50% because of the pulse discontinuity. Halving the maximum possible DAC output voltage lowers the DAC's signal-to-noise ratio which reduces the overall sensitivity of the setup.

The digitization of the PA signal is done in two operation modes depending on the microphone's clocking signal: the standard mode (detecting signals between 100 Hz and 10 kHz) and the ultrasonic mode (detecting signals between 10 kHz and 80 kHz). Switching between modes of operation requires changing of the microphone's clocking frequency. Therefore, continuous measurements between 8 kHz to 62 kHz are not possible and have to be divided in two parts.

### 2.2 Simulation models

This section briefly describes how the two models are realized and implemented. Ideally, the simulation model would reflect the measured geometry (explained in section 2.1.1) since the PA signal depends on the exact details of the resonator's geometry. However, this would considerably increase the computation demands of the model and the simulation time. Therefore, an average diameter of each cylinder was calculated as seen in Figure 2 and used as the cylinder's diameter during simulations.

The design of the resonator is initially generated and meshed. We applied swept meshes in the resonance cylinder and parts of the cavity cylinder, while a triangular mesh was used in areas of the resonator where a swept mesh could not be used. A structured mesh with swept meshing was chosen to reduce the number of mesh elements and enable faster computation. Even though the AME model does not require boundary layers, they were generated throughout the resonator since the same mesh was used for the VT model. A convergence study performed for the AME model indicated that the mesh depicted in Figure 4 is sufficient to obtain a good approximation to the actual solution.

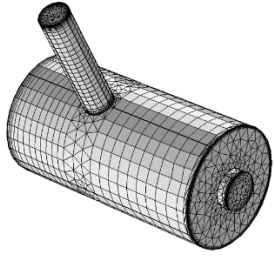

**Figure 4:** The resonator mesh (22,451 elements).

### 2.2.1 Amplitude mode expansion model (AME)

The AME method is based on eigenmode expansion

$$p(\mathbf{r},\omega) = \sum_j A_j(\omega) p_j(\mathbf{r}) \qquad (1)$$

where $p(\mathbf{r},\omega)$ is the acoustic pressure at the measurement point $\mathbf{r}$ and modulation frequency $\omega$. It is executed using a MATLAB® code which accesses COMSOL Multiphysics® for calculations of the eigenmodes $p_j(\mathbf{r})$ and the eigenfrequencies $\omega_j$ of the resonator by solving the acoustic Helmholtz equation using a sound hard boundary condition. The amplitudes $A_j(\omega)$ are calculated using

$$A_j(\omega) = i \frac{\mathcal{A}_j \omega}{\omega^2 - \omega_j^2 + i\omega\omega_j l_j}. \qquad (2)$$

$\mathcal{A}_j$ describes the excitation of the sound waves and is obtained by

$$\mathcal{A}_j = \frac{\alpha(\gamma-1)}{V_C} \int_{V_C} p_j^* I \, dV, \qquad (3)$$

where $V_C$ is volume of the resonator and $I = I(\mathbf{r})$ is the laser intensity profile in the sample where the absorbing molecules are located. $\gamma$ is the ratio of isobaric and isochoric heat capacity and $\alpha$ is the absorption coefficient of the sample. The loss effects are introduced by loss factors $l_j$ in Equation (2), which account for thermal and viscous losses at the bulk of the fluid and at the resonator's surface, hindering $A_j(\omega) \to \infty$ as $\omega \to \omega_j$; in fact, due to the imaginary unit $i$, the amplitude always remains bounded. Detailed description of the formulas can be found elsewhere [11,15]. Air is selected as the propagating fluid with the parameters in Table 1.

**Table 1**: Air parameters at a temperature of 20°C and a static pressure of 1013 hPa [13].

| | |
|---|---|
| Density | 1.2044 kg/m³ |
| Sound velocity | 343.2 m/s |
| Viscosity | 1.814 10⁻⁵ Pa s |
| Coefficient of heat conduction | 2.58 10⁻² W/m K |
| Specific heat capacity at constant volume | 7.1816 10² J/kg K |
| Specific heat capacity at constant pressure | 1.0054 10³ J/kg K |

The PA signal was simulated between 8 to 62 kHz with an increment of 10 Hz. The source term for sound generation is defined within the absorption cylinder as shown in Figure 5. Small variations of the shape and the size of the heat source have no significant effect on the simulation results [15].

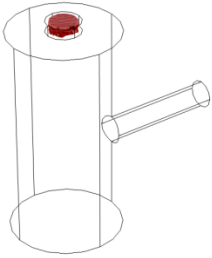

**Figure 5**: The position of the source term (red) within the resonator.

### 2.2.2 Viscothermal model (VT)

The VT method is based on solving the Navier-Stokes equation, the continuity equation for the mass, and the energy balance equation. An equation of state is introduced to relate the variations in pressure, temperature and density. A detailed description of the equations can be found elsewhere [12]. The model is simulated using COMSOL Multiphysics® software.

The walls of the PA cell are set as sound hard (no slip and isothermal boundary conditions). Air was selected from the COMSOL Multiphysics® material database as the propagating fluid in the resonator. The air properties like the dynamic viscosity, thermal conductivity, heat capacity at constant pressure and the density are temperature dependent. The temperature was set to 20°C and the static pressure to 1013 hPa. The source term was defined just like in the AME model.

The VT model had more variables than the AME model, hence it had 476,980 degrees of freedom while the AME model had 115,474. This is the reason why the VT model is considerably slower and computationally more demanding. The PA signal of the VT model was calculated from 8 to 62 kHz. Due to the long computing times the frequency increment was set to 50 Hz.

### 3. Results

As previously explained, the microphone has two operating modes and continuous measurements over the complete frequency range are not possible. The simulated and measured frequency response curves between 10 kHz and 62 kHz are shown in Figure 6. The experimental data was smoothed using the Savitzky Golay function in Matlab [17] and rescaled to enable easy comparison. The measurement results between 8 kHz and 10 kHz where, according to the simulations, the first resonance peak is expected are presented in Figure 7.

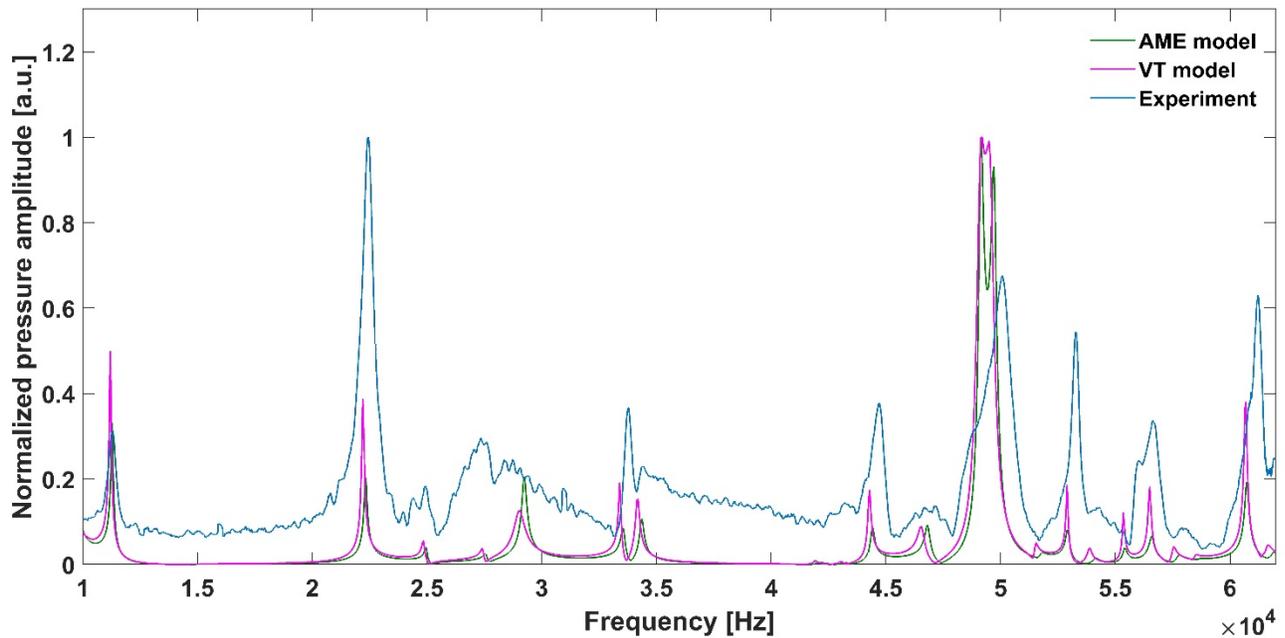

**Figure 6:** The frequency responses of AME model, VT model and measurement in the high frequency range.

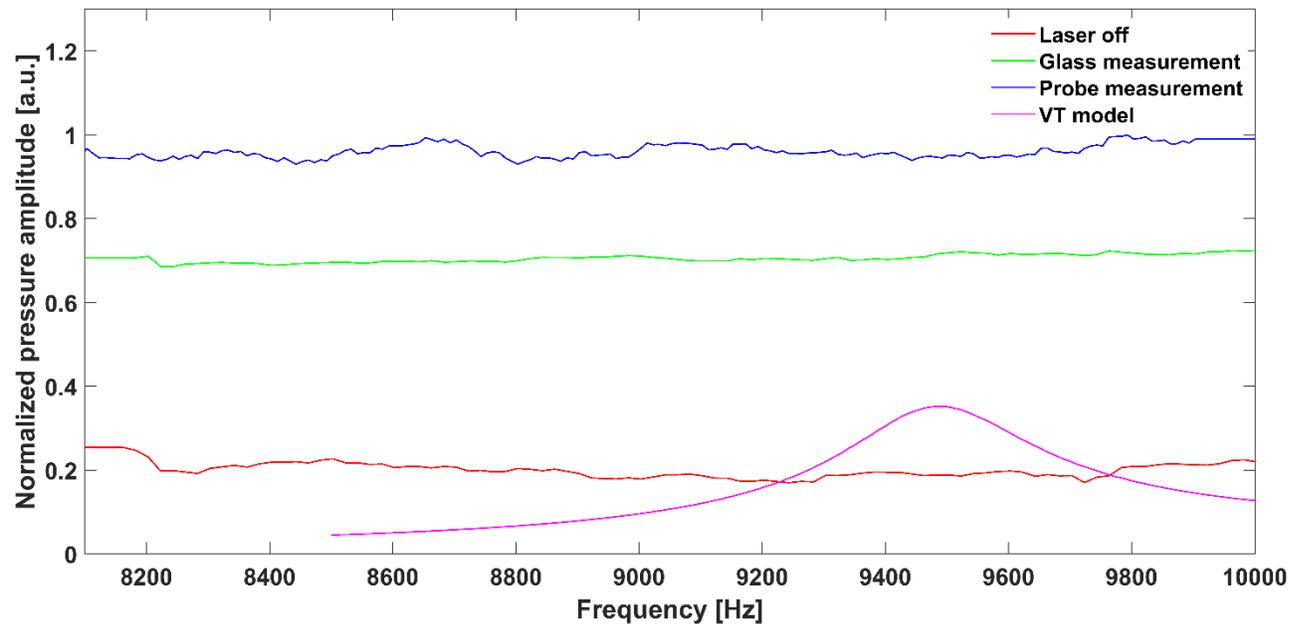

**Figure 7:** Experimental frequency response in the low frequency range. The red plot is the offset level when the laser is turned off, the green plot is the measurement with the glass window while the blue plot is with the graphite probe.

## 4. Discussion
### 4.1. Measurement

The measurement is compared with the VT model since it is considered the more accurate simulation method. The resonance frequencies from the measurement show fairly good accordance with the VT simulations. The deviation is not more than 1.1%. However, the measured resonances are broader than the simulation resonances. Table 2 compares the $Q$ factors of the most prominent resonances. It indicates a considerably higher damping in the measurement system compared to the numerical model. The $Q$ factor was estimated using:

$$Q = \frac{f_0}{f_h - f_l} \tag{4}$$

where $f_o$ denotes the resonance frequency while $f_l$ and $f_h$ are the frequencies at which the value of the pressure amplitude has decreased to half of the resonance value. Due to the high damping, the measured $Q$ values are obtained from resonances where the $Q$ factor could be estimated.

**Table 2:** Resonances $Q$ factors of measurement and simulation (see Figure 6).

| $f_{res}$ in kHz | $Q_{sim}$ | $Q_{meas}$ |
|---|---|---|
| 11.2 | 80 | 19 |
| 22.2 | 123 | 34 |
| 44.3 | 211 | 55 |
| 52.9 | 353 | 124 |
| 60.7 | 304 | 80 |

The damping can be attributed to leakage of the photoacoustic signal from the resonator. We suspect that the microphone mount does not seal the resonator tightly and could be the location of signal leakage. Furthermore, the resonator's walls have a rough surface as indicated earlier, which increases viscous losses at the surfaces. In cylindrical cells with a length to diameter ratio significantly larger than one, the surface roughness is important for longitudinal modes since the sound particle velocity is along the cylinder barrel [15]. The base noise level of the experimental response function is comparably high. This might be caused by laser excitation of the resonator's walls. The measurement system currently employs a diode laser which characteristically has an elliptical divergent beam and it is suspected that it might be inducing a PA signal of the cell walls. In addition, laser excitation of the PA cell's window could be contributing to the noise level since the resonator does not have buffer zones.

Almost all the resonances, which appear in the response curves of the simulations, have been reproduced by the experimental measurement. The simulation model predicts two resonances between 30 kHz and 40 kHz. However, only a single resonance is measured in this region. The absence of one of the peaks can possibly be attributed to the fact that the resonance could not be distinguished from the base noise level which is highest between 35 kHz and 40 kHz. The simulation model predicts 2 distinguishable resonances between 55 kHz and 57 kHz. However, broadening of the measured resonances has resulted in the merging of the two resonances.

Unlike the simulation, the measured response curve in Figure 7 does not clearly show a resonance between 8 kHz and 10 kHz. The signal difference between the laser off measurement (red) to the glass measurement (green) supports our assumption that there is window and cell wall excitation.

### 4.2. Comparison of the simulation models

Both simulations were carried out with the same computer. The simulation time for the VT model was 1 week while the AME model took 7 minutes. All the VT model resonances are reproduced by the AME model with a deviation of less than 1.8%. The broad resonance peak between 48 kHz and 51 kHz is a result of two overlapping resonance peaks at 49.2 kHz and 49.5 kHz. At low modulation frequencies the resonances are attributed mainly to longitudinal modes of the cavity cylinder. At frequencies above 50 kHz, radial modes are also supported by the resonator thus accounting for the increase in the number of resonances (Table 3).

Since sensing applications ask for a strong PA signal, we restrict the following discussion to the 14 resonances with amplitudes that exceed a relative value of $10^{-6}$. It can be observed that the relative height of the resonance amplitudes for both models depends on the primary location of the mode within the resonator. If the mode is mainly located in the resonance cylinder where the surface area to volume ratio is large, the AME model underestimates the losses in comparison to the VT model which results in larger amplitudes than the latter (ratio of AME to VT model resonance amplitude > 1).

If the mode is mainly located in the cavity cylinder, where the surface area to volume ratio is small, the AME model slightly overestimates the losses in comparison to the VT model (resonance amplitude ratio < 1). In some cases, the mode occupies the cavity cylinder as well as the resonance cylinder. Then the amplitude ratio can be smaller or larger than 1. The relation between surface area to volume ratio and loss leads to the conclusion, that the AME model tends to overestimate bulk loss effects and underestimates the surface loss effects in comparison to the VT model. In the case of modes spanning both cylinders, the two effects mitigate each other.

**Table 3**: Resonance frequency, corresponding mode, location of strong antinodes and amplitude ratio for the 14 strongest resonances. The location of the antinodes has been determined by lifting the lower limit of the depicted data range appropriately.

| $f_{\text{res}}$ in kHz | $\|p\|$-profile of acoustic mode | Main location of the mode | $\dfrac{A^{\text{AME}}(\omega_{\text{res}})}{A^{\text{VT}}(\omega_{\text{res}})}$ |
|---|---|---|---|
| 9.500 | 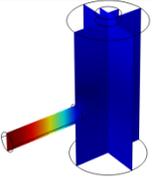 | 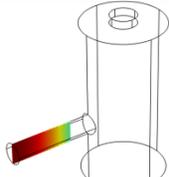 Resonance cylinder | 2.43 |
| 11.200 | 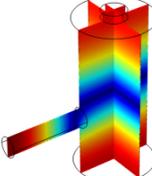 | 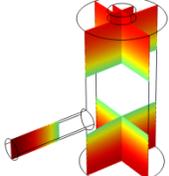 Resonance and cavity cylinder | 0.91 |
| 22.200 | 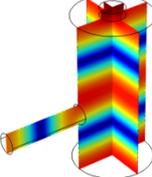 | 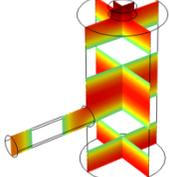 Resonance and cavity cylinder | 0.72 |
| 29.050 | 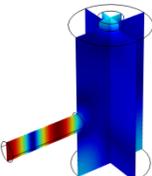 | 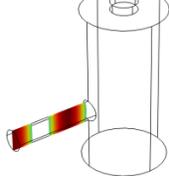 Resonance cylinder | 2.23 |
| 33.400 | 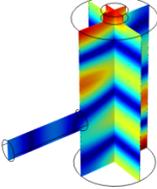 | 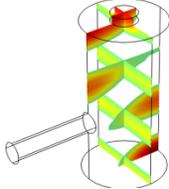
Cavity cylinder | 0.59 |
| 34.200 | 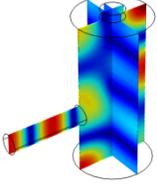 | 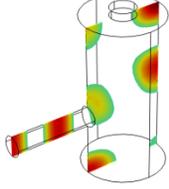
Resonance and cavity cylinder | 0.97 |
| 44.300 | 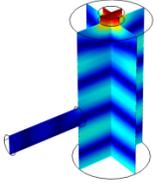 | 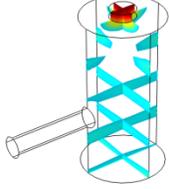
Cavity cylinder | 0.68 |
| 46.550 | 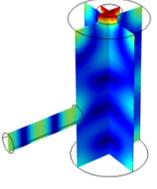 | 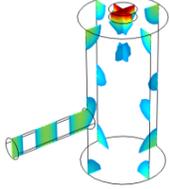
Resonance and cavity cylinder | 1.42 |
| 49.200 | 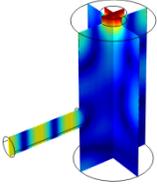 | 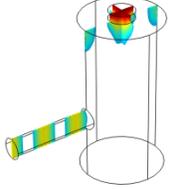
Resonance and cavity cylinder | 1.30 |
| 49.500 | 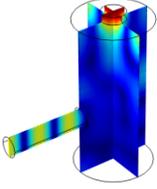 | 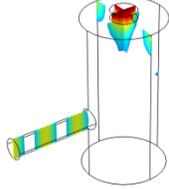
Resonance and cavity cylinder | 1.27 |

| | | | |
|---|---|---|---|
| 52.900 | 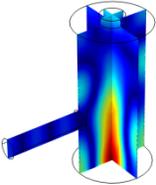 | 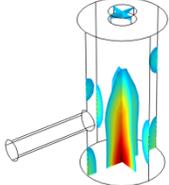<br>Cavity cylinder | 0.60 |
| 55.350 | 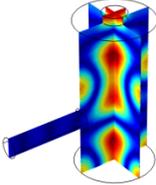 | 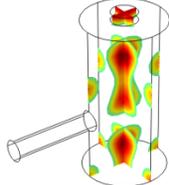<br>Cavity cylinder | 0.47 |
| 56.500 | 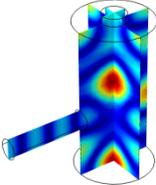 | 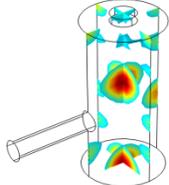<br>Cavity cylinder | 0.49 |
| 60.700 | 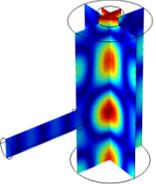 | 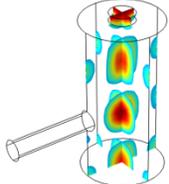<br>Cavity cylinder | 0.65 |

**5. Outlook and conclusion**

We have shown the ability of the AME method to accurately simulate the PA signal in T-cell resonator. The method provides a much quicker and computationally less demanding alternative to the VT method for photoacoustic simulations in macroscopic resonators. The measured resonances and simulated resonances showed fairly good accordance. The width of the measured resonances indicates the need for additional loss mechanisms to be implemented in the modelling.